\documentclass[english]{article}
\usepackage[T1]{fontenc}
\usepackage[latin9]{inputenc}
\usepackage{wrapfig}
\usepackage{amstext}
\usepackage{graphicx}
\usepackage[numbers]{natbib}

\makeatletter

\providecommand{\tabularnewline}{\\}

\usepackage{microtype}
\usepackage{graphicx}
\usepackage{subfigure}
\usepackage{booktabs} % for professional tables
\usepackage{resizegather}

\usepackage{hyperref}

\usepackage[nonatbib,final]{neurips_2020}

\author{
  \normalfont Aaron Defazio \hspace*{1em} Tullie Murrell \\
  Facebook AI Research, \\
  Facebook New York \\
  \and
  Michael P. Recht \\
  Department of Radiology \\
  NYU Grossman School of Medicine \\
}

\makeatother

\usepackage{babel}
\begin{document}
\title{MRI Banding Removal via Adversarial Training}
\maketitle
\begin{abstract}
MR images reconstructed from sub-sampled Cartesian data using deep
learning techniques show a characteristic banding (sometimes described
as streaking), which is particularly strong in low signal-to-noise
regions of the reconstructed image. These unnatural artifacts have
been identified as one of the largest obstacles preventing the clinical
use of machine-learning based MRI reconstructions. In this work, we
propose the use of an adversarial loss that penalizes banding structures
without requiring any human annotation. Our technique greatly reduces
the appearance of banding, without requiring any additional computation
or post-processing at reconstruction time. Our approach is compatible
with any existing reconstruction approach that uses supervised machine
learning, including the current state-of-the-art. We report the results
of a blind comparison against a strong baseline by a group of expert
evaluators (board-certified radiologists), where our approach is ranked
superior at banding removal with no statistically significant loss
of detail. A reference implementation of our method is available in
the supplementary material.
\end{abstract}

\section{Introduction}

The use of deep-learning approaches for accelerating MR imaging has
recently shown significant promise \citep{Hammernik2018,Schlemper2018,aggarwal2017},
with learning approaches far out-performing classical penalized least
squares approaches for reconstructing images from raw subsampled k-space
signals. However, existing approaches produce images with some unnatural
structures that prevent radiologists from accepting the images for
clinical use despite the advantages the images have over classical
techniques with respect to reconstruction accuracy metrics such as
the structured similarity metric (SSIM).\begin{wrapfigure}{o}{0.5\columnwidth}%
\begin{centering}
\includegraphics[width=2.6in]{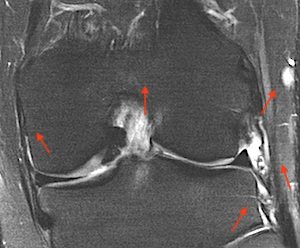}
\par\end{centering}
\caption{\label{fig:banding-example}A deep-learning MRI reconstruction that
shows significant horizontal banding artifacts}
\vspace{-3em}
\end{wrapfigure}%

In this work, we describe a method for removing the primary artifact
produced by Cartesian deep-learning reconstruction systems: banding.
This banding, as illustrated in Figure \ref{fig:banding-example},
is characterized by a streaking pattern exactly aligned with the phase-encoding
direction (horizontal in the figure). This banding is anisotropic
and non-homogenous across the image, and most visible in high-noise
or low-contrast areas. It is the result of the signal subsampling
process used during Cartesian accelerated MRI, whereby subsampling
occurs in one spatial direction only. 

At the 2019 \textit{Medical Imaging meets NeurIPS} workshop \citep{knoll2020advancing},
banding artifacts were shown to occur in reconstructions from each
of the top 3 winning teams in the competition, despite the significant
differences between the reconstruction approaches. These artifacts
were identified as a major obstacle to the use of machine-learning
based reconstructions in clinical practice. 

Our paper is structured as follows: In Section \ref{sec:accel} we
formally describe the MRI reconstruction problem as it applies to
current clinical MRI scanners. In Section \ref{sec:adversary} we
describe how to augment standard deep-learning based MRI reconstruction
methods with our orientation adversary. Section \ref{sec:models}
describes the state-of-the-art reconstruction model that we use in
our experiments. In Section \ref{sec:masking} we describe the masking
procedure we use, in Section \ref{sec:baselines} we describe the
classical baseline that we compare against, and in Section \ref{sec:training}
we detail how our model was trained. Finally in Section \ref{sec:evaluation}
we detail the results of a blind evaluation by radiologists.

\section{Accelerated Parallel 2D MRI}

\label{sec:accel}In MR imaging, a spatial image is produced by combining
measurements of the anatomy acquired in the Fourier domain, known
as k-space. Let $m$ be the true greyscale spatial image. Classical
approaches produce an image of the anatomy by acquiring a full cartesian
grid of samples from k-space, then applying the inverse fast Fourier
transform. In our notation the estimated greyscale image $\hat{m}$
is:
\[
\hat{m}=\mathcal{F}^{-1}(x)
\]
where $x$ is a $h\times w$ matrix of k-space measurements. In this
work, we focus on 2D MRI images produced by a modern MR imaging system,
which contain two additional complications: parallelization and acceleration.

\subsection{Parallel Imaging}

In a parallel MRI system, more than one receiver coil is used, resulting
in a tensor of acquired k-space images. Instead of each coil imaging
the entire field-of-view, each coil covers a smaller portion of the
anatomy. The signal acquired by coil $i$ of $n_{c}$ coils is given
by a Fourier transform $\mathcal{F}$:
\begin{equation}
x_{i}=\mathcal{F}(s_{i}\circ m)+\text{noise},\label{eq:parallel-eq}
\end{equation}
where $s$ is a complex-valued coil-sensitivity map that is applied
element-wise. The $s_{i}$ values can be estimated via well-known
auto-calibration procedures, but this is not necessary for all deep-learning
approaches.

The coil signals are commonly combined using the \textit{root-sum-squares}
(RSS) procedure to produce a spatial image. The RSS estimate at pixel
$l,m$ is given by:
\[
m_{\text{RSS},lm}=\sqrt{\sum_{i=1}^{n_{c}}\left|m_{i,lm}\right|^{2}},
\]
where for each coil $m_{i}=\mathcal{F}^{-1}(x_{i})$ is the individual
coil image. The RSS estimate produces images with slightly (often
negligibly) higher noise than more sophisticated approaches but has
the advantage of robustness and simplicity.

\subsection{Accelerated imaging}

\vspace{-0.2em}
Accelerated MRI systems capture a subset of the full k-space system
to reduce scan time. If parallel imaging is used, the system of linear
equations given by Equation \ref{eq:parallel-eq} is overdetermined
even when a subset of the $x_{i}$ pixel values are known as long
as the sub-sampling factor is less than the number of coils, and so
a least-squares solve may be used to produce a spatial image, at the
expense of an increase in noise over non-accelerated MRI \citep{grappa,sense}.
Two-fold subsampling is widely used in clinical practice, typically
using either the SENSE \citep{sense} or GRAPPA \citep{grappa} classical
approaches to reconstruction. Higher acceleration factors can be achieved
using regularized least-squares in the case of sparse anatomy such
as vascular MRI \citep{sparsemri}, but these approaches produce poor
results for general-purpose MR imaging.

\subsection{Machine learning approaches to accelerated 2D MRI}

In the machine learning approach to 2D MRI reconstruction, a training
set of $n_{\text{data }}$ instances (slices) is gathered, where each
instance is a $k$-space tensor $x^{(j)}:n_{c}\times h\times w$.
Then standard parallel imaging is used to produce spatial images $m^{(j)}$.
A scan of a patient consists of multiple spatially consecutive slices
with different scan modalities of the same anatomy, although for training
purposes we treat the slices independently and sample from the total
set of slices across all patients in the dataset i.i.d.

These training pairs are then used to train a black box predictor
$B_{\phi}$, which maps from raw subsampled k-space tensors (complex
valued multichannel Fourier domain images) to greyscale $h\times w$
spatial images. Given an image-space loss function $l$, the training
loss for datapoint $j$ is:
\[
L^{(j)}(\phi)=l\left(B_{\phi}\left(M\left(x^{(j)}\right)\right),\,m^{(j)}\right)
\]
where $M$ is a masking function that zeros out a fraction of the
$k$-space lines. We detail this masking function further in the Section
\ref{sec:masking}. Essentially, the model $B$ is trained to produce
an image as close as possible to the ``ground-truth'' fully-sampled
parallel MRI as possible, using only a fraction of the data following
the standard empirical risk minimization setup. In our experiments,
we used a combination of SSIM and L1 losses with weighting 0.01 for
the L1 component.

\section{Orientation Adversary}

\begin{figure}[t]
\begin{centering}
\includegraphics[clip,width=1\textwidth]{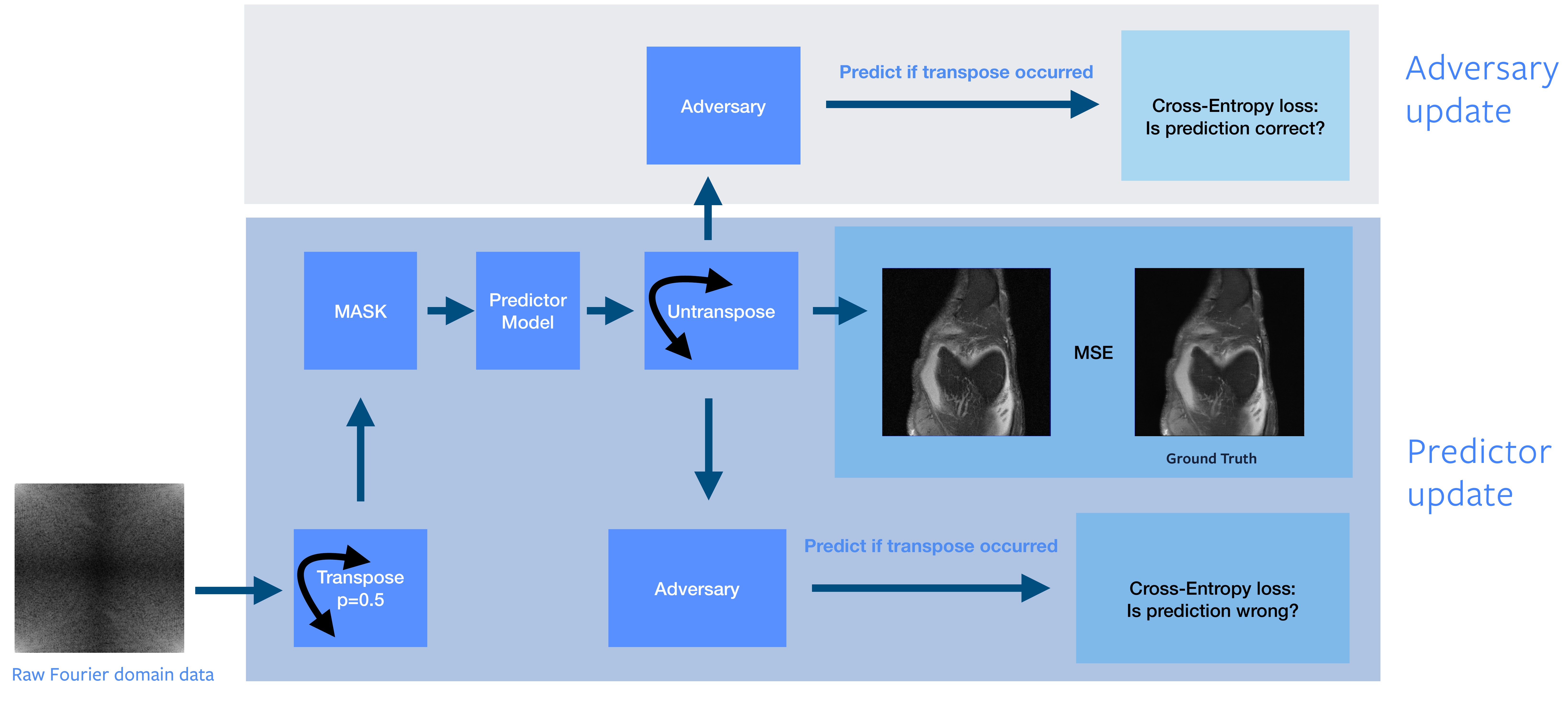}
\par\end{centering}
\centering{}\caption{Orientation adversarial training}
\end{figure}
\label{sec:adversary}The primary difficulty with removing banding
is the lack of available annotation; direct supervised machine learning
techniques can not be used. Our insight is to use image orientation
during reconstruction as a self-supervised learning signal. Image
banding is aligned with the direction of subsampling in the mask,
in our case horizontally. This signal can be used by an adversarial
training term to identify and penalize banding.

In particular, let $A_{\theta}$ be an adversary model, which maps
from spatial images $h\times w$ to real values $[0,1]$. Its goal
is to predict if the given spatial image contains horizontal $(0$)
or vertical $(1)$ banding. We train this adversary simultaneously
with the predictor model, using the same minibatch to compute stochastic
gradient steps simultaneously for both, rather than in an alternating
fashion.

\paragraph{Predictor}

The training of our prediction model is modified as follows. Before
applying $B$, a random operator is sampled using a Bernoulli variable
$r^{(j)}$ with probability 0.5, either a random flip $R_{1}$ (transpose
$h\longleftrightarrow w$) or the identity operator $R_{0}$. This
operator is then applied before and after the application of $B$:
\[
\hat{m}^{(j)}=\left(R_{r^{(j)}}\circ B_{\phi}\circ M\circ R_{r^{(j)}}\right)x^{(j)}.
\]
The predictor's loss is then augmented with a term that encourages
it to produce images that fool the adversary:
\[
L_{B}^{(j)}(\phi)=l\left(\hat{m}^{(j)},\,m^{(j)}\right)+\text{CE}\left(1-r^{(j)},A_{\theta}\left(\hat{m}^{(j)}\right)\right),
\]
where $\text{CE}$ is the binary cross-entropy. Gradients are not
propagated to the adversaries parameters $\theta$ during the predictor
update, but are fully propagated through the adversary from output
to input then through the predictor to its parameters $\phi$. This
is accomplished by toggling requires\_grad\_ to false for the adversaries
parameters in Pytorch during loss calculation, then toggling it back
before the adversary loss calculation.

\paragraph{Adversary}

The adversary is trained to predict the $r$ variable directly, with
the addition of a regularization term:
\[
L_{A}^{(j)}(\theta)=\text{CE}\left(r^{(j)},A_{\theta}\left(\hat{m}^{(j)}\right)\right)+\gamma\left\Vert \nabla_{\hat{m}^{(j)}}A_{\theta}(\hat{m}^{(j)})\right\Vert ^{2}.
\]
Gradients are not propagated through the predictor during the adversary
step, using Pytorch's detach operator on $\hat{m}^{(j)}$. We found
that regularization of the adversary is necessary for stable convergence.
We use a simplified gradient penalty as used by \citep{gradient-penalty},
which is closely related to WGAN-GP penalty \citep{wgangp} which
is widely used for Generative Adversarial Network (GAN) training. 

\section{Models}

\label{sec:models}Our technique is applicable using any predictor
and adversary model architecture. We detail the two models we used
below. Full source code is available in the supplementary material. 

\paragraph*{Predictor}

We used the state-of-the-art predictor architecture from \citep{varnet2}
consisting of a sequence of 12 U-Net models, interleaved with a soft-projection
onto the known k-space lines. This sequence is followed by an inverse
Fourier transform then a root-sum-squares operation to produce the
final image. Each U-Net has 12 channels after the first convolution
and has 4 pooling operations (i.e. convolutions occur at 5 resolutions
when the input resolution is included). This network has only 13 million
parameters, but more than 250 convolutional layers. This model is
the current state-of-the-art for MRI reconstruction.

\paragraph{Adversary}

The shallow ResNet consists of an initial convolution to increase
the channel count to 64, followed by 2 pre-activation ResNet basic
blocks, 4x4 max-pooling, 2 more blocks with 128 channels, then 4x4
max-pooling followed by average pooling down to a 1x1 image, a ReLU
then a linear layer. This architecture was not heavily optimized,
and we expect a better architecture could give even better results.
We found initially that a ResNet-50 architecture took too long to
converge, and that it was necessary to include ResNet blocks that
apply at the full-resolution of the image, rather than after a downsampling
operation as performed in the standard ResNet architecture. Because
of the small batch-sizes used for training MRI reconstruction models,
we found that it was necessary to replace BatchNorm with GroupNorm
\citep{groupnorm}, and we used group-size 32 as our default.

\section{Masking}

\label{sec:masking}In 2D Cartesian MR imaging, the two directions
in the Fourier domain are known as the Frequency and Phase encoding
directions. The most common clinical practice for accelerated MRI
is to only capture every second frequency-encoding line, except for
a band of the lowest-frequency lines which are always included. These
lines are typically stored electrically with zero values filling the
unseen values. All coil images for a slice follow the same mask as
the lines are captured simultaneously during acquisition. The acquisition
process can rapidly acquire an entire line in the frequency direction
line at once, so typically no subsampling is performed within the
lines.

For the fastMRI dataset, the phase-encoding direction is always horizontal,
and we use a central band of 16 low-frequency lines. These lines are
only consecutive in k-space if the Fourier domain is viewed with the
low-frequency lines in the center of the image, which is NOT the standard
layout returned by fast-Fourier routines. The low-frequency lines
serve multiple purposes \citep{grappa,sense}:
\begin{itemize}
\item They allow estimation of the coil sensitivities through auto-calibration, 
\item they contain a significant fraction of the total signal energy. 
\item They allow easier disambiguation of aliasing produced by the mask.
\end{itemize}
Machine learning models trained using the same number of total lines
but without a central region do not produce images of as high quality.
The disambiguation function is important. A direct IFFT of a masked
k-space coil image produces two images of the anatomy, with one shifted
by half the image width. This is known as aliasing. Without additional
k-space lines, the signal does not contain information indicating
which of the two positions in the image the anatomy is in. The central
low-frequency k-space lines contain information allowing a perfect
reconstruction of a low-resolution version of the full coil image,
which the machine learning model can use to disambiguate the two positions.

For our accelerated reconstruction we performed 4-fold acceleration
by sampling every 4th frequency line, using an offset equispaced mask
following \citep{defazio2019offset}, with the central lines as discussed
above. This level of acceleration produces significant artifacts when
used with classical accelerated MRI techniques, but produces very
high-quality results when using deep-learning reconstructions. We
also experimented with the random masks preferred by the theory of
compressed sensing; however, we found they performed significantly
worse. This is consistent with prior work on machine learning reconstructions
\citep{varnet2,defazio2019offset}. We attribute this to the lack
of sparsity required by compressed sensing theory for the Proton Density
(PD) and PD Fat Saturated (PDFS) MRI images used in our experiments.

\section{Baselines}

\begin{figure*}
\begin{centering}
\includegraphics[width=0.9\textwidth]{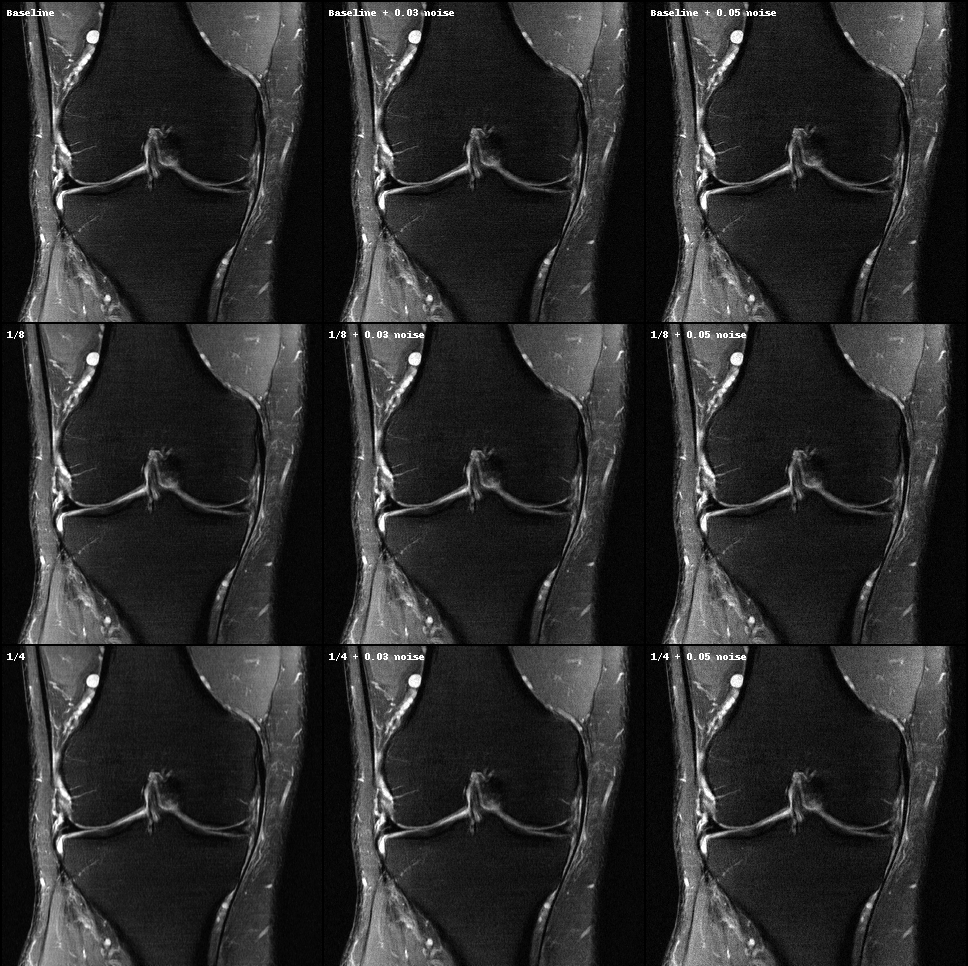}
\par\end{centering}
\caption{\label{fig:blurring}Baseline (top), with varying levels of blurring,
against our adversarial method (bottom). Columns show different levels
of adaptive noise. Images are best viewed on a high-brightness monitor
at full resolution.}
\end{figure*}
\label{sec:baselines}As the banding problem is relatively new we
are not aware of any existing baseline methods for banding removal
specific to MR imaging. We will instead apply a classical image processing
method for minimizing the visibility of banding artifacts in images
known as dithering \citep{floyd1975,jarvis1976}. Dithering is simply
the process of adding noise in a specific way to improve the perceptual
quality of an image. We found that adding noise directly to the reconstruction
improved apparent sharpness, but did not by itself reduce the visibility
of banding (Figure \ref{fig:blurring}). However, if an anisotropic
blurring filter is used before the addition of noise, banding was
well-suppressed. We used the following convolutional kernel parametrized
by $\alpha$:
\[
K=\frac{1}{1+2\alpha}\left[\begin{array}{ccc}
0 & \alpha & 0\\
0 & 1 & 0\\
0 & \alpha & 0
\end{array}\right],
\]
which blurs in the vertical direction only, the opposite direction
from the banding pattern. Figure \ref{fig:blurring} shows that some
regions are banding free for $\alpha=1/8$, but banding suppression
improves as blurring is further increased. The addition of noise is
crucial, the blurred images in the left column for $\alpha=1/8$ and
$1/4$ still show signs of banding that are much less apparent after
noise is applied (middle and right column). The noise also increases
perceptual sharpness, so the usage of blurring is no longer immediately
apparent. The noise used here is Gaussian noise with variance equal
to the median of the image in an $11\times11$ pixel region, multiplied
by a constant. 

The dithering method has the clear disadvantage of both decreasing
the effective resolution of the image and decreasing the signal-to-noise
ratio. This is clear when examining fine detail in Figure \ref{fig:blurring}.
The higher of the two noise levels shown (0.05) clearly masks banding
much more significantly than a noise level of 0.03, however it was
determined to be too much noise to be used clinically after consultation
with a radiologist. We settled on a noise level of 0.03 as a level
which is high but still a clinically relevant baseline.

Another more subtle problem arises from the inhomogeneity of the amount
of banding across the image. Applying a uniform amount of dithered
across the image results in some regions not receiving enough dithering,
and other regions that showed no banding being unnecessarily dithered.
Ideally, the dithering would be applied adaptively across the image,
in proportion to the amount of banding; however, this requires a model
or method that outputs the ``banding amount'' in a region, which
is a non-trivial problem. Our proposed adversarial method implicitly
learns such a model, as part of the discriminator term in its loss.

Since dithering can improve the perceptual image quality as assessed
by a viewer while at the same time reducing signal-to-noise, resolution,
and measures such as MSE and SSIM compared to the ground truth image,
its use must be carefully considered. We determined that blurring
up to $\alpha=1/8$ may be of use clinically but that $\alpha=1/4$
losses too much fine detail.

\section{Training}

\label{sec:training}We trained our models on the knee scans from
the fastMRI dataset \citep{zbontar2018fastMRI} using the approach
detailed above. This is the first large-scale dataset released of
raw full-sampled k-space. For the evaluation, we trained and evaluated
on scans from 1.5 Tesla machines only as banding is less visible in
scans from 3 Tesla machines making evaluation more difficult, although
our approach works equally well for 3T scans. Training consisted of
100 pre-training epochs, where the adversarial term is not used (but
the flip operator is included), using ADAM with learning rate 0.0003,
momentum 0.9 and batch-size 8 (1 per GPU on an 8-gpu system) and no
weight decay. The pre-training was followed by 60 epochs of training
including the adversarial term, with learning rate 0.0001, and adversary
gradient regularization strength $\gamma=0.1$. We trained with a
factor of 4 acceleration, using 16 central k-space lines, using the
train/test splits distributed as part of the fastMRI dataset.

\section{Evaluation by Radiologists}

\label{sec:evaluation}We set up our reader study as a three-fold
comparison against the adversarial method, the non-adversarial method,
and the dithering baseline. The non-adversarial method, which uses
an identical setup to the adversarial approach excepting the adversarial
terms in the loss, will be denoted the ``standard'' approach. Each
of the six board-certified radiologists in our study were given a
set of 20 of 40 volumes from the validation set (Each volume is approximately
25 slice images), so that each volume of the 40 was evaluated three
times independently. 

Each radiologist was asked to rank the three methods in terms of the
degree of banding and separately in terms of the retention of fine
detail (the questionnaire given to each radiologist is in the Supplementary
Material) for the volumes provided to them. Equal adaptive noise as
detailed in Section \ref{sec:baselines} was used on all three to
avoid the known bias of human evaluators to assess noisy images as
having higher detail levels.

They were given access to a ground-truth corresponding to the non-accelerated
(fully sampled) images. For each volume, the methods were assigned
designations ``A'', ''B'', ``C'', randomized at the volume level,
to ensure that the evaluators were blind to the reconstruction method
used. A two-sided paired sign-test with Bonferroni multiple comparison
correction to the p-values was used. The tests were performed using
the average rank from the 3 radiologist who ranked each volume. Volume
evaluations are considered independent for the purposes of the test,
and the pairing is at the within-volume level. The statistical analysis
was chosen at the experimental design stage to avoid statistical fishing.
The raw experimental results and R script used for our analysis are
given in the Supplementary Material.

\begin{table*}
\begin{centering}
\begin{tabular}{|c|c|c|c|}
\hline 
p-values & Standard & Dithering & Average rank (higher is better)\tabularnewline
\hline 
\hline 
Adversary & $1.09\times10^{-11}$ & $2.18\times10^{-11}$ & \textbf{2.83}\tabularnewline
\hline 
Dithering & 0.028 & - & 1.74\tabularnewline
\hline 
Standard & - & - & 1.43\tabularnewline
\hline 
\end{tabular}\vspace{-0.3em}
\caption{\texttt{\textbf{Banding Removal }}-- results of the two-sided pairwise
comparison with Bonferroni correction for the following question proposed
to board-certified radiologists: ``\textbf{Rank the 3 methods in
terms of the amount of visible banding}''.\protect \\
Our proposed method is ranked as better than the two baselines with
very high statistical significance.}
\vspace{0.5em}
\par\end{centering}
\begin{centering}
\begin{tabular}{|c|c|c|c|}
\hline 
p-values & Standard & Dithering & Average rank (higher is better)\tabularnewline
\hline 
\hline 
Adversary & 2.61 & $8.82\times10^{-4}$ & \textbf{2.18}\tabularnewline
\hline 
Dithering & $3.25\times10^{-6}$ & - & 1.5\tabularnewline
\hline 
Standard & - & - & \textbf{2.32}\tabularnewline
\hline 
\end{tabular}\vspace{-0.3em}
\caption{\texttt{\textbf{Detail retention }}-- results of the two-sided pairwise
comparison with Bonferroni correction for the following question proposed
to board-certified radiologists: ``\textbf{Rank the 3 methods in
terms of the retention of fine anatomical detail in comparison to
the ground truth}''. Our proposed method is not statistically significantly
different in terms of detail retention from the standard baseline,
and highly statistically significantly better than the dithering approach.
P-values range from 0 to 3 due to the Bonferroni correction.}
\vspace{0.5em}
\par\end{centering}
\centering{}%
\begin{tabular}{|c|c|c|}
\hline 
Adversary & Standard & Dithering\tabularnewline
\hline 
\hline 
\textbf{0.725} (0.62-0.82) & 0.967 (0.92-1) & 0.975 (0.90-0.99)\tabularnewline
\hline 
\end{tabular}\vspace{-0.3em}
\caption{\texttt{\textbf{Presence of banding }}-- two-sided 95\% binomial
confidence intervals with Bonferroni correction for the question proposed
to board-certified radiologists ``\textbf{Is any banding present
(yes, no)}''. Our approach completely removes all traces of banding
27.5\% of the time. The standard and dithering approaches are rarely
reported as having no banding by radiologists (<4\% of the time).\vspace{-1em}
}
\end{table*}

\subsection{Evaluation Results}

The radiologists ranked our adversarial approach as better than the
standard and dithering approaches with an average rank of 2.83 out
of a possible 3. This result is statistically significantly better
than either alternative with p-values $1.09\times10^{-11}$ and $2.18\times10^{-11}$
respectively, and the adversarial approach was ranked as the best
or tied for best in 85.8\% of 120 total evaluations (95\% CI: 0.78-0.91).
The dithering approach is also statistically significantly better
than the standard approach.

We also asked radiologists if banding was present (in any form) in
the reconstructions in each case. This evaluation is highly subjective,
as ``banding'' is hard to define in a precise enough way to ensure
consistency between evaluators. Considering each radiologist's evaluation
independently, on average banding is still reported to be present
in 72.5\% (\textbf{95\% }CI: 0.62-0.82) of cases even with the adversarial
learning penalty. The radiologists were not consistent in their rankings;
the overall percentages reported by the six radiologists were 20\%,
75\%, 75\%, 80\%, 85\%, and 100\% for the adversarial reconstructions.
In contrast, for the baseline and dithered reconstructions, only one
radiologist reported less than 100\% presence of banding for each
method (80\% and 85\% presence respectively, from different radiologists). 

\begin{figure*}
\begin{centering}
\hspace{1.2cm}Ground Truth\hspace{3.2cm}Standard \hspace{2.5cm}Orientation
Adversary\hfill{}
\par\end{centering}
\begin{centering}
\includegraphics[width=1\textwidth]{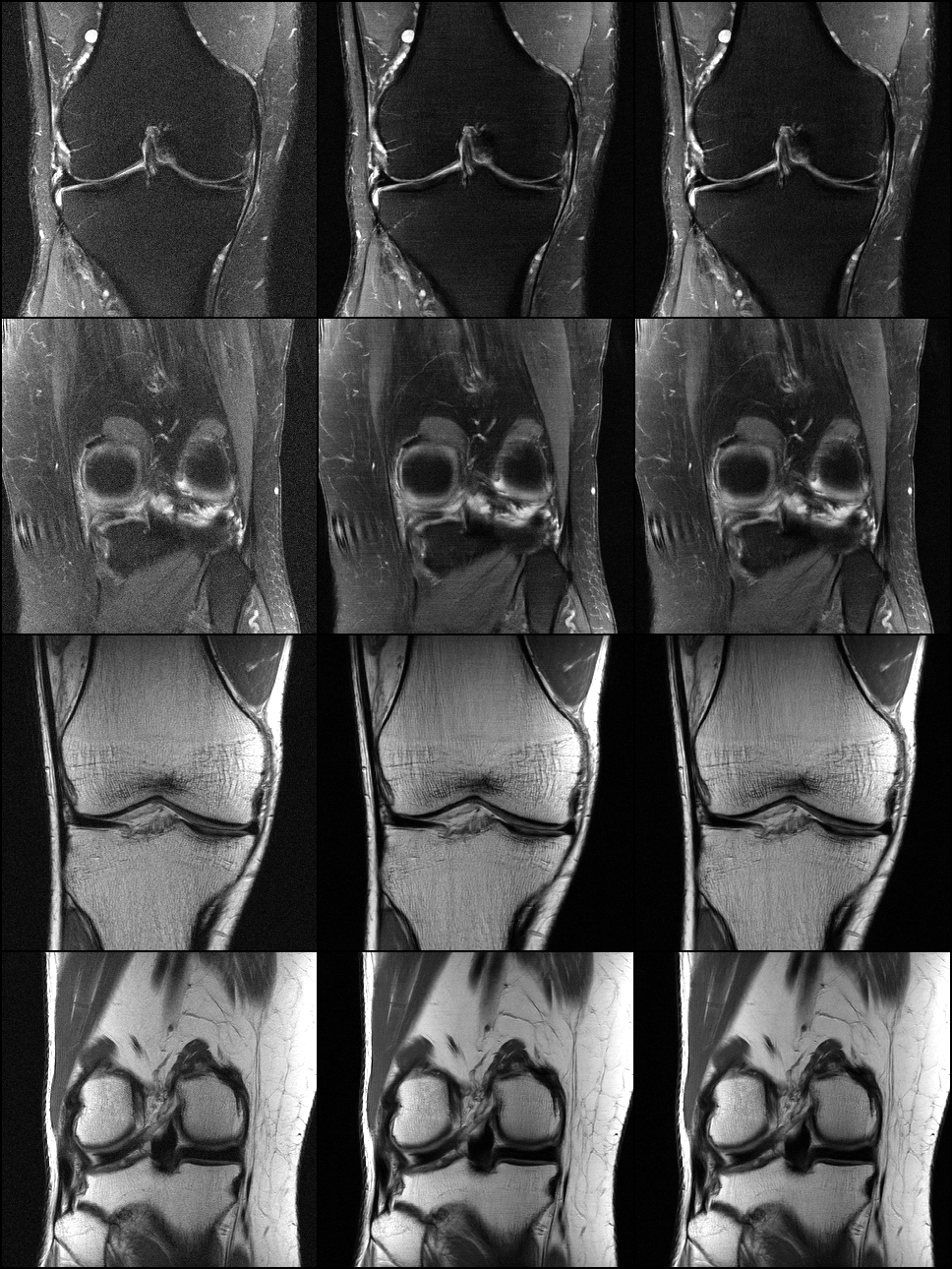}
\par\end{centering}
\centering{}\caption{\label{fig:examples}A comparison of ground-truth images against a
standard accelerated deep-learning reconstruction and the proposed
orientation adversary. Two fat-suppressed and two non-fat suppressed
images are shown, chosen at random from the validation set. Images
are best viewed on a high-brightness monitor.}
\end{figure*}
\nopagebreak

\section{Related Work}

\vspace{-0.5em}
Machine-learning based reconstruction of MRI images is an ongoing
research direction and not currently in clinical use, so very little
discussion of the downsides to current techniques exists in the literature.
We are not are of any work on banding removal of accelerated MRI images;
however, other forms of banding introduced by balanced steady state
free precession (bSSFP) sequences have been investigated, using both
physics-based \citep{signal-banding} and supervised machine learning
\citep{nnbandingremoval} approaches. These techniques are not directly
applicable to the removal of acceleration banding artifacts.

Adversarial learning has been applied to medical imaging in a number
of recent works, for instance for segmentation \citep{advsegmentation2017},
Magnetic Resonance Angiography (MRA) image generation \citep{Olut2019},
super-resolution \citep{gan-superresolution2019}, and generation
of anonymized datasets \citep{artifical-mri}. Adversarial training
has also been applied directly for MRI reconstruction \citep{gan-cs2019}.
We believe caution should be used when applying adversarial penalties
to directly encourage reconstructions to resemble non-accelerated
ground-truth images, they are prone to introducing phantom anatomical
detail. Our use of an adversarial penalty to detect orientation does
not suffer from this problem as it only ever takes reconstructed images
as input rather than ground-truth images.

\section*{Conclusion}

\vspace{-0.5em}
In this work, we have presented an effective technique for producing
machine learning models for accelerated MRI that minimize visible
banding artifacts. Our technique is \textbf{not} specifically tied
to any reconstruction approach, it may be be applied on top of any
MRI reconstruction technique that uses supervised machine learning.

\section*{Acknowledgements}

This work was made possible through close collaboration with the fastMRI
team at Facebook AI research, including Nafissa Yakubova, Anuroop
Sriram, Jure Zbontar, Larry Zitnick, Mark Tygert and Suvrat Bhooshan,
and our fastMRI project members at NYU Langone Health \citep{zbontar2018fastMRI},
with special thanks to Florian Knoll, Matthew Muckley and Daniel Sodickson.

\section*{Broader Impact Statement}

The use of machine learning in medical imaging raises a number of
ethical and societal considerations. The benefits to the use of our
method are clear; images produced using our method have fewer image
artifacts than standard reconstructions. Our method, used in combination
with current machine-learning based reconstruction methods, has the
potential to significantly reduce time spent in MRI scanners, and
the associated cost of scanning. This benefits both patients and medical
professionals.

In theory, the removal of the banding artifacts may lead to the occlusion
of fine anatomical detail that would otherwise be present and useful
for diagnosis. Our evaluations by radiologists are strong evidence
that the method does not remove detail, but we believe a larger study
directly aimed at determining differences in diagnosis would be required
to establish this. It's not a-priori certain that any method of banding
removal will reduce detail, since the ground truth does not have banding,
the use of a banding removal method is just a use of prior knowledge
in the Bayesian sense.

Machine learning approaches to MRI reconstruction are potentially
prone to biases in the training data. If anatomy outside of the norms
of the training data is encountered, the reconstruction may not be
accurate. When using our proposed approach, the same considerations
are necessary as for any application of machine learning in medical
imaging. At a minimum the model must be tested for robustness to outlier
examples.

\bibliographystyle{plain}
\bibliography{banding_removal}

\begin{thebibliography}{10}

\bibitem{aggarwal2017}
Hemant Aggarwal, Merry Mani, and Mathews Jacob.
\newblock Modl: Model based deep learning architecture for inverse problems.
\newblock {\em IEEE Transactions on Medical Imaging}, PP, 12 2017.

\bibitem{signal-banding}
Marcus Bj{\"o}rk, Erik Gudmundson, Jo{\"e}lle~K. Barral, and Petre Stoica.
\newblock Signal processing algorithms for removing banding artifacts in mri.
\newblock {\em Proceedings of the 19th European Signal Processing Conference
  (EUSIPCO-2011)}, 2011.

\bibitem{defazio2019offset}
Aaron Defazio.
\newblock Offset sampling improves deep learning based accelerated mri
  reconstructions by exploiting symmetry.
\newblock Technical report, Facebook, 2019.

\bibitem{floyd1975}
R.~Floyd and L.~S. Steinberg.
\newblock An adaptive algorithm for spatial gray scale.
\newblock In {\em International Symposium Digest of Technical Papers}, 1975.

\bibitem{grappa}
MA~Griswold, PM~Jakob, RM~Heidemann, M~Nittka, V~Jellus, J~Wang, B~Kiefer, and
  A~Haase.
\newblock Generalized autocalibrating partially parallel acquisitions (grappa).
\newblock {\em Magnetic Resonance in Medicine}, 2002.

\bibitem{wgangp}
Ishaan Gulrajani, Faruk Ahmed, Martin Arjovsky, Vincent Dumoulin, and Aaron
  Courville.
\newblock Improved training of wasserstein gans.
\newblock In {\em Proceedings of the 31st International Conference on Neural
  Information Processing Systems (NeurIPS 2017)}, 2017.

\bibitem{Hammernik2018}
Kerstin Hammernik, Teresa Klatzer, Erich Kobler, Michael~P. Recht, Daniel~K.
  Sodickson, Thomas Pock, and Florian Knoll.
\newblock Learning a variational network for reconstruction of accelerated mri
  data.
\newblock {\em Magnetic Resonance in Medicine}, 79(6):3055--3071, 2018.

\bibitem{jarvis1976}
J.F. Jarvis, C.N. Judice, and W.H. Ninke.
\newblock A survey of techniques for the display of continuous tone pictures on
  bilevel displays.
\newblock {\em Computer Graphics and Image Processing}, 1976.

\bibitem{artifical-mri}
Koshino Kazuhiro, Rudolf Werner, Fujio Toriumi, Mehrbod Javadi, Martin Pomper,
  Lilja Solnes, Franco Verde, Takahiro Higuchi, and Steven Rowe.
\newblock Generative adversarial networks for the creation of realistic
  artificial brain magnetic resonance images.
\newblock {\em Tomography (Ann Arbor, Mich.)}, 12 2018.

\bibitem{nnbandingremoval}
Ki~Hwan Kim and Sung-Hong Park.
\newblock Artificial neural network for suppression of banding artifacts in
  balanced steady-state free precession mri.
\newblock {\em Magnetic Resonance Imaging}, 37:139 -- 146, 2017.

\bibitem{knoll2020advancing}
Florian Knoll, Tullie Murrell, Anuroop Sriram, Nafissa Yakubova, Jure Zbontar,
  Michael Rabbat, Aaron Defazio, Matthew~J. Muckley, Daniel~K. Sodickson,
  C.~Lawrence Zitnick, and Michael~P. Recht.
\newblock Advancing machine learning for mr image reconstruction with an open
  competition: Overview of the 2019 fastmri challenge, 2020.

\bibitem{sparsemri}
Michael Lustig, David Donoho, and John~M. Pauly.
\newblock Sparse mri: The application of compressed sensing for rapid mr
  imaging.
\newblock {\em Magnetic Resonance in Medicine}, 2007.

\bibitem{gan-superresolution2019}
Qing Lyu, Chenyu You, Hongming Shan, Yi~Zhang, and Ge~Wang.
\newblock {Super-resolution MRI and CT through GAN-CIRCLE}.
\newblock In {\em Developments in X-Ray Tomography XII}. International Society
  for Optics and Photonics, 2019.

\bibitem{gan-cs2019}
M.~{Mardani}, E.~{Gong}, J.~Y. {Cheng}, S.~S. {Vasanawala}, G.~{Zaharchuk},
  L.~{Xing}, and J.~M. {Pauly}.
\newblock Deep generative adversarial neural networks for compressive sensing
  mri.
\newblock {\em IEEE Transactions on Medical Imaging}, 2019.

\bibitem{gradient-penalty}
Lars Mescheder, Andreas Geiger, and Sebastian Nowozin.
\newblock Which training methods for {GAN}s do actually converge?
\newblock In {\em Proceedings of the 35th International Conference on Machine
  Learning}, 2018.

\bibitem{advsegmentation2017}
Pim Moeskops, Mitko Veta, Maxime~W. Lafarge, Koen A.~J. Eppenhof, and Josien
  P.~W. Pluim.
\newblock Adversarial training and dilated convolutions for brain mri
  segmentation.
\newblock In {\em Deep Learning in Medical Image Analysis and Multimodal
  Learning for Clinical Decision Support}, Cham, 2017. Springer International
  Publishing.

\bibitem{Olut2019}
Sahin Olut, Yusuf~H. Sahin, Ugur Demir, and Gozde Unal.
\newblock Generative adversarial training for mra image synthesis using
  multi-contrast mri.
\newblock In {\em PRedictive Intelligence in MEdicine}, Cham, 2018. Springer
  International Publishing.

\bibitem{sense}
K~P Pruessmann, M~Weiger, M~B Scheidegger, and P~Boesiger.
\newblock Sense: sensitivity encoding for fast mri.
\newblock {\em Magnetic Resonance in Medicine}, 1999.

\bibitem{Schlemper2018}
J.~{Schlemper}, J.~{Caballero}, J.~V. {Hajnal}, A.~N. {Price}, and
  D.~{Rueckert}.
\newblock A deep cascade of convolutional neural networks for dynamic mr image
  reconstruction.
\newblock {\em IEEE Transactions on Medical Imaging}, 2018.

\bibitem{varnet2}
Anuroop Sriram, Jure Zbontar, Tullie Murrell, Aaron Defazio, C.~Lawrence
  Zitnick, Nafissa Yakubova, Florian Knoll, and Patricia Johnson.
\newblock End-to-end variational networks for accelerated mri reconstruction.
\newblock Technical report, Facebook, 2019.

\bibitem{groupnorm}
Yuxin Wu and Kaiming He.
\newblock Group normalization.
\newblock {\em International Journal of Computer Vision}, 2019.

\bibitem{zbontar2018fastMRI}
Jure Zbontar, Florian Knoll, Anuroop Sriram, Matthew~J. Muckley, Mary Bruno,
  Aaron Defazio, Marc Parente, Krzysztof~J. Geras, Joe Katsnelson, Hersh
  Chandarana, Zizhao Zhang, Michal Drozdzal, Adriana Romero, Michael Rabbat,
  Pascal Vincent, James Pinkerton, Duo Wang, Nafissa Yakubova, Erich Owens,
  C.~Lawrence Zitnick, Michael~P. Recht, Daniel~K. Sodickson, and Yvonne~W.
  Lui.
\newblock {fastMRI}: An open dataset and benchmarks for accelerated {MRI}.
\newblock 2018.

\end{thebibliography}

\end{document}